# Shannon Entropy Quantum Entanglement Criterion and the Generalized Uncertainty Principle

**Otto Gadea, Gardo Blado**[*]

*Department of Mathematics and Physics, College of Science and Mathematics*

*Houston Baptist University*

*7502 Fondren Rd., Houston, Texas, U.S.A*

**Abstract**

We examine quantum gravity effects by applying the generalized uncertainty principle (GUP) to entropic uncertainty relation conditions on quantum entanglement. In particular, we study the GUP corrections to the Shannon entropic uncertainty condition for entanglement. As in an earlier paper [1] which dealt with variance relations, there is an increase in the upper bound for the entanglement condition upon the application of the generalized uncertainty principle. Fundamental concepts of the generalized uncertainty principle, entanglement and the entropic uncertainty relations are also discussed.

[*] Corresponding author:
*Email address*: gblado@hbu.edu

# 1. Introduction

The idea of a minimal length scale which limits the distance at which one can probe nature has been motivated by thought experiments involving gravitational effects in quantum mechanics and various attempts to formulate a quantum theory of gravity (such as string theory, loop quantum gravity, non-commutative geometry, etc.). String theory and certain thought experiments have been shown to give rise to a generalized uncertainty principle (GUP) from which models are developed which result in a minimal length scale [2]. In this paper, we will use the most widely used model which is through the modified commutation relations.

The Heisenberg Uncertainty Principle (HUP) of ordinary quantum mechanics is given by (variance form)

*Equation 1:* $\Delta x \Delta p \geq \frac{1}{2}$ ($\hbar = 1$).

This is derivable from the commutator of $x$ and $p$, $[x, p] = i$, using [3] $(\Delta A)^2 (\Delta B)^2 \geq \left(\frac{1}{2i}\langle[\hat{A},\hat{B}]\rangle\right)^2$ with $A = x$ and $B = p$. String scattering has been shown to give rise to a modification of the HUP to a GUP [2, 4-6]. In this paper, we will use

*Equation 2:* $\Delta x \Delta k \geq \frac{1}{2}(1 + \beta(\Delta k)^2)$

as the form of the GUP which results from the modified commutation relation $[x, k] = i(1 + \beta k^2)$. $k$ here is the momentum operator representation in terms of a GUP and $\beta$ is the small positive GUP correction parameter. It is related to the HUP momentum operator by [7]

*Equation 3:* $k = \frac{\tan(\sqrt{\beta} p)}{\sqrt{\beta}}$.

Equation 2 exhibits a non-zero minimal length [8]. There is a rich literature which discuss the GUP to which we refer the reader [2, 8-9]. Our short introduction above is meant to establish notation and discuss the most relevant equation for the discussions below.
Similar to [1], we study in the present paper, entanglement in continuous variable systems but instead of the variance form in Equation 1, we will make use of entanglement conditions using the Shannon entropic uncertainty relations. The paper is organized as follows. In section 2 we give an elementary discussion of entanglement in terms of a finite dimensional bipartite system of qubits and briefly state how entanglement and the HUP can be related for a continuous variable system. Section 3 involves an account of the entropic uncertainty relations (Shannon entropy in particular). We work out in detail the Shannon entropic quantum entanglement criterion for a bipartite system involving EPR-type operators in section 4 to make the derivation of the GUP correction apparent in section 5. Some conclusions are given in section 6.

## 2. Entanglement and the HUP

Particles are said to be entangled if their quantum states are somehow linked such that knowing the state of one particle determines the states of the others. Let us give an example using a qubit bipartite system to make this more explicit[1].

Suppose we have a system of two electrons (1 and 2). Let the state of electron 1 be

*Equation 4:* $|\psi_1\rangle = \alpha_1|0\rangle + \beta_1|1\rangle$

and electron 2 be

*Equation 5:* $|\psi_2\rangle = \alpha_2|0\rangle + \beta_2|1\rangle$.

The composite (pure) state of the system can be written as a tensor product (indicated by the operator $\otimes$), $|\psi\rangle = |\psi_1\rangle \otimes |\psi_2\rangle = (\alpha_1|0\rangle + \beta_1|1\rangle) \otimes (\alpha_2|0\rangle + \beta_2|1\rangle) = \alpha_1\alpha_2|0\rangle\otimes|0\rangle + \alpha_1\beta_2|0\rangle\otimes|1\rangle + \alpha_2\beta_1|1\rangle\otimes|0\rangle + \beta_1\beta_2|1\rangle\otimes|1\rangle$. A convenient notation is to let $|A\rangle\otimes|B\rangle = |AB\rangle$. So we can write the preceding equation as

*Equation 6:* $|\psi\rangle = \alpha_1\alpha_2|00\rangle + \alpha_1\beta_2|01\rangle + \alpha_2\beta_1|10\rangle + \beta_1\beta_2|11\rangle$.

Note that the expansion coefficients are such that $|\alpha_1\alpha_2|^2 + |\alpha_1\beta_2|^2 + |\alpha_2\beta_1|^2 + |\beta_1\beta_2|^2 = 1$.

Next, consider a state

*Equation 7:* $|\psi\rangle = \frac{1}{2\sqrt{2}}|00\rangle - \frac{\sqrt{3}}{2\sqrt{2}}|01\rangle + \frac{1}{2\sqrt{2}}|10\rangle - \frac{\sqrt{3}}{2\sqrt{2}}|11\rangle$

which is known to be the composite system made up of 2 states $|\psi_1\rangle$ and $|\psi_2\rangle$ of Equation 4 and Equation 5. We can factorize this into 2 states.

*Equation 8:* $|\psi\rangle = \frac{1}{\sqrt{2}}|0\rangle\left(\frac{1}{2}|0\rangle - \frac{\sqrt{3}}{2}|1\rangle\right) + \frac{1}{\sqrt{2}}|1\rangle\left(\frac{1}{2}|0\rangle - \frac{\sqrt{3}}{2}|1\rangle\right) = \left(\frac{1}{\sqrt{2}}|0\rangle + \frac{1}{\sqrt{2}}|1\rangle\right) \otimes \left(\frac{1}{2}|0\rangle - \frac{\sqrt{3}}{2}|1\rangle\right) = |\psi_1\rangle \otimes |\psi_2\rangle$

with with $|\psi_1\rangle = \left(\frac{1}{\sqrt{2}}|0\rangle + \frac{1}{\sqrt{2}}|1\rangle\right)$ and $|\psi_2\rangle = \left(\frac{1}{2}|0\rangle - \frac{\sqrt{3}}{2}|1\rangle\right)$. Factorizable states are separable states [11]. Hence we say that $|\psi\rangle$ in Equation 7 is a separable state.

However, given a composite state, it is not always possible to factorize it into individual states! For example, consider the state,

---
[1] We will follow the discussion in [10].

*Equation 9:* $|\psi'\rangle = \frac{1}{\sqrt{2}}|00\rangle + \frac{1}{\sqrt{2}}|11\rangle$.

If this is to factorize, then the right hand side of Equation 9 must be equal to the tensor product of two states. Let these two general states be as in Equation 4 and Equation 5. From Equation 9 and Equation 6, $\frac{1}{\sqrt{2}}|00\rangle + \frac{1}{\sqrt{2}}|11\rangle = \alpha_1\alpha_2|00\rangle + \alpha_1\beta_2|01\rangle + \alpha_2\beta_1|10\rangle + \beta_1\beta_2|11\rangle$. Comparing coefficients, we get $\alpha_1\alpha_2 = \frac{1}{\sqrt{2}}$ so $\alpha_1$ and $\alpha_2$ CANNOT EITHER BE ZERO. We also get $\beta_1\beta_2 = \frac{1}{\sqrt{2}}$ so $\beta_1$ and $\beta_2$ CANNOT EITHER BE ZERO. However we also get $\alpha_1\beta_2 = 0$ which means that either $\alpha_1 = 0$ or $\beta_2 = 0$!!! We have a contradiction! Hence there is no solution and we cannot write $|\psi\rangle$ of Equation 9 as a tensor product of the two states in Equation 4 and Equation 5. $|\psi'\rangle = \frac{1}{\sqrt{2}}|00\rangle + \frac{1}{\sqrt{2}}|11\rangle$ is an example of an <u>entangled state</u>. We can say that for the entangled state

*Equation 10:* $|\psi'\rangle \neq |\psi_1\rangle \otimes |\psi_2\rangle$.

In the discussion above we can consider the state $|0\rangle$ to describe a particle with spin UP and $|1\rangle$ a particle with spin DOWN. So for composite systems, the state $|10\rangle$ will describe the composite state in which particle 1 is spin down and particle 2 is spin up, etc. If we measure the spin of particle 1 in Equation 9 to be up then we are sure that the second particle will have spin up also because of the composite state $|00\rangle$. This, however is not true for Equation 7. Due to the presence of the two terms $|00\rangle$ and $|01\rangle$, if particle 1 is spin up $|0\rangle$, there is a probability that the second particle is also spin down $|1\rangle$! Hence Equation 9 is called an <u>entangled state</u> because if we know the state of one particle, we surely know the state of the other unlike in Equation 7.

The above example which makes use of a finite dimensional system, is a standard starting point in discussing the concept of entanglement. To explore the connection of the HUP and entanglement, we need to look at continuous variable systems which will involve the position and momentum operators. An excellent introductory discussion on entanglement beyond spin systems can be found in [12]. A more advanced treatment is given in [13].

Similar to Equation 10 an entangled composite state represented by $\Psi(x_1, x_2)$ in coordinate space or $\wp(p_1, p_2)$ in momentum space is such that $\Psi(x_1, x_2) \neq \psi_1(x_1)\psi_2(x_2)$ or $\wp(p_1, p_2) \neq \varphi_1(p_1)\varphi_2(p_2)$ where $\psi_i(x_i)$ $(\varphi_i(p_i))$ could represent the coordinate (momentum) wave function of particle *i*. With the states dependent on the position and momentum (which are continuous variables), and with the uncertainty principle expressed in terms of the position and momentum operators in Equation 1, once could draw connections between the entanglement and the HUP [14-17] and subsequently with the GUP [1].

# 3. Entropic Uncertainty Relations

For two observables that do not commute, such as position and momentum, Equation 1 addresses a fundamental limit to the extent with which one can know about one of the particle's properties if the other has already been measured. For years, the favored approach towards uncertainty in quantum mechanics was the use of standard deviations, in Equation 1 primarily because in probability and statistics, the dispersion is the most well-known gauge of uncertainty in measurements [18].

However, as pointed out by [18], it was not the ideal measure, indicated by two major flaws that appeared in the Robertson relation [19], the expression from which Equation 1 can be derived when some non-commuting observables $A$ and $B$ have been specified to be position and momentum:

*Equation 11:* $\Delta A \Delta B \geq \frac{1}{2}|\langle \Psi|[\hat{A},\hat{B}]|\Psi\rangle|$

The first weakness is that in a finite, N-dimensional Hilbert space—a generalized version of the usual Euclidean "3D" space—the lower bound in the right-hand side becomes dependent upon the state of the particle, which becomes especially problematic when the state is an eigenvalue of the observable. Equation 11 yields zero which basically eliminates the restriction on the variances of $A$ and $B$. The second weakness is that for conjugate variables with continuous probability distributions, the dispersion may not be the useful measure of uncertainty especially if the distributions have several peaks. This is especially true for applications in information theory where standard deviation does not play an important role [20].

To overcome these issues, alternative uncertainty relations were sought out by physicists. Recently, an approach utilizing entropic uncertainty relations has been gaining traction due to the relations' clear validity as mathematical inequalities [21]. Consequently, it has been postulated that such relations may bear physical significance in measurements and so the search of a useful relation, if any exist, for analysis of quantum mechanical systems has been the topic of debate [22-23]. The desirable features of the entropic uncertainty relations over the HUP have been discussed in [20].

In favor of the view that entropic uncertainty relations can be useful in experiments, many physicists attempt to explain what entropy can mean in a quantum context. In one particular definition, entropy can be seen as the amount of information one lacks when one witnesses an event that could have many outcomes [24]. This information can be used to characterize its source based not on what the information is about, but on how much of it is transmitted compared to how much of it can be transmitted.

One way to look at it is similar to Shannon's own approach about compressing and decompressing messages in which some information cannot be compressed and then sent, so it is lost.

Recall the difference between one's thoughts and one's speech. A person can think about something with all their knowledge about it at their disposal, but expressing what they think into words may not convey everything that they know to a listener. The lack of a complete view about what the speaker knows about something, would be reflected in information entropy just like the measurement of some particle's wavefunction.

To understand the definition of the particular entropic function we will be using, let us first go back to the Boltzmann entropy $S = k_B \ln W$, of classical thermodynamics and statistical mechanics [25-26] where $k_B$ is the Boltzmann's constant and $W$ is the number of microscopic configurations that a system can be in, assuming equiprobable configurations. Hence $W = 1/P$ where $P$ is the probability of each configuration. Thus $S/k_B = -\ln P$. However if the different configurations are not equiprobable, then we need to add the weight-factor of them occurring as we sum over the different probabilities $P_i$,

*Equation 12:* $S/k_B = -\sum_i P_i \ln P_i$.

From Equation 12, Shannon [24] defined the (dimensionless) average information entropy content of a random variable X (called the Shannon entropy) as

*Equation 13:* $H \equiv \sum_x P_X(x) \ln(1/P_X(x)) = -\sum_x P_X(x) \ln P_X(x)$,

where $P_X(x)$ is the probability distribution of the random variable X to have the measured value $x$. Note that if the probability of measuring X as x is 1 then there is no information entropy or no uncertainty. The lower the probability of measuring x, the higher the information entropy or uncertainty. Hence information entropy can be a considered a measure of uncertainty.

Extending Equation 13 to continuous variables, we get the differential Shannon entropy (or we will refer to this as simply "Shannon entropy" in the succeeding sections) as

*Equation 14:* $H = -\int_{x_1}^{x_2} P(x) \ln(P(x)) dx$

where $P(x)$ is the probability distribution of the continuous variable $x$.

It is apparent that we can define the Shannon entropy involving position and momentum probability distributions from Equation 14 as

*Equation 15:* $H[w] = -\int_{-\infty}^{\infty} dx\, w(x) \ln w(x);\ H[v] = -\int_{-\infty}^{\infty} dp\, v(p) \ln v(p)$

where $w(x) \equiv |\psi(x)|^2$ and $v(p) \equiv |\varphi(p)|^2$. Bialynicki-Birula and Mycielski [27] derived an entropic uncertainty relation ("BBM inequality") involving the Shannon entropy similar to the HUP.

*Equation 16:* $H[w] + H[v] \geq \ln(\pi e)$

It has been shown that HUP follows from Equation 16. The above expression owes its validity to a fact that Heisenberg himself acknowledged—to gain information about the position of a particle, one must lose information about the momentum of that same particle and vice versa due to their nature as conjugate variables [28]. Essentially, as the entropy of one property decreases, the entropy of the other must increase to maintain the total uncertainty of the particle above the limit shown in the right-hand side of Equation 16. In addition, although the HUP describes uncertainty due to a lack of the simultaneous existence of precise values for both position and momentum, it ignores the uncertainty induced by the measurement of an actual instrument, which Equation 16 can account for [21]. Coles, et. al [20] discussed how Equation 16 overcomes the two weaknesses of the HUP discussed above.

Since the BBM inequality is a reliable representation of the uncertainty inherent in knowing the properties of one particle, it will also be useful when considering two particles which apparently involves more information. This forms the basis of an entropic uncertainty relation that can be used to detect a specific characteristic of the two-particle system, namely entanglement, due to the fact that the measurement of one correlated particle results with information on the other particle as well, decreasing the total uncertainty of the system beyond its expected limit. Identification of entangled particles has always proven inconvenient with factorization since the process can sometimes be tedious and it is difficult to be certain that some composite states are definitively inseparable. A lack of factorizability can indicate that particles are entangled, but it can also mean that the composite state is simply very complex, so an entropic criterion that can guarantee entanglement would be very useful indeed.

## 4. Shannon Entropy Quantum Entanglement Criterion

The following discussion will follow closely Walborn, et.al [29]. Considering a bipartite system, we define operators similar to the annihilation operators defined in [30], which are linear combinations of the position and momentum operators $x_j$ and $p_j$ with $j = 1, 2$, pertaining to subsystems 1 and 2, and the usual commutation relation, $[x_j, p_k] = i\delta_{jk}$.

*Equation 17:* $r_j = \cos\theta_j x_j + \sin\theta_j p_j$ , $s_j = \cos\theta_j p_j - \sin\theta_j x_j$ .

From Equation 17, we define further the operators, $r_\pm = r_1 \pm r_2$ and $s_\pm = s_1 \pm s_2$. As in [31], we study EPR-type operators by letting $\theta_1 = \theta_2 = 0$ in Equation 17 which give, similar to the general $r_\pm$ and $s_\pm$ operators,

*Equation 18:* $x_\pm = x_1 \pm x_2$ ; $p_\pm = p_1 \pm p_2$ .

Consider a separable pure state $|\Psi\rangle = |\boldsymbol{\psi_1}\rangle \otimes |\boldsymbol{\psi_2}\rangle$ with the corresponding wavefunction in coordinate space, $\Psi(x_1, x_2) = \boldsymbol{\psi_1}(x_1)\boldsymbol{\psi_2}(x_2)$ and in momentum space $\wp(p_1, p_2) = \varphi_1(p_1)\varphi_2(p_2)$. A change in variables in Equation 18, yields

*Equation 19:* $\frac{1}{\sqrt{2}}\psi_1\left(\frac{x_++x_-}{2}\right)\psi_2\left(\frac{x_+-x_-}{2}\right) = \Psi(x_+,x_-)$ ; $\frac{1}{\sqrt{2}}\varphi_1\left(\frac{p_++p_-}{2}\right)\varphi_2\left(\frac{p_+-p_-}{2}\right) = \wp(p_+,p_-)$.

Let us compute the Shannon entropy associated with the measurement of $x_\pm$ and $p_\pm$. From the definition of the Shannon entropy in Equation 15,

$H[w_\pm] = -\int_{-\infty}^{\infty} dx_\pm w_\pm(x_\pm)\ln w_\pm(x_\pm)$; $H[v_\pm] = -\int_{-\infty}^{\infty} dp_\pm v_\pm(p_\pm)\ln v_\pm(p_\pm)$, where $w_\pm(x_\pm) = w_\pm(x_1,x_2)$ and $v_\pm(p_\pm) = v_\pm(p_1,p_2)$ are the probability distributions of $x_\pm$ and $p_\pm$ respectively. From Equation 19, these are given by

*Equation 20:* $w_\pm(x_\pm) = \int dx_\mp |\Psi(x_+,x_-)|^2 = \frac{1}{2}\int dx_\mp\ |\psi_1|^2|\psi_2|^2$ ; $v_\pm(p_\pm) = \int dp_\mp |\wp(p_+,p_-)|^2 = \frac{1}{2}\int dp_\mp\ |\varphi_1|^2|\varphi_2|^2$.

Let us define $w_i(x_i) \equiv |\psi_i(x_i)|^2$ and $v_i(p_i) \equiv |\varphi_i(p_i)|^2$. Hence Equation 20 becomes

*Equation 21:* $w_\pm(x_\pm) = \frac{1}{2}\int dx_\mp\ w_1\left(\frac{x_++x_-}{2}\right)w_2\left(\frac{x_+-x_-}{2}\right)$ and $v_\pm(p_\pm) = \frac{1}{2}\int dp_\mp\ v_1\left(\frac{p_++p_-}{2}\right)v_2\left(\frac{p_+-p_-}{2}\right)$.

For convenience, as will be apparent later, we make a further change in variables in terms of $\frac{x_++x_-}{2} = x_1 = x$ and so $\frac{x_+-x_-}{2} = \mp x \pm x_\pm$ and similarly $\frac{p_++p_-}{2} = p_1 = p$ and so $\frac{p_+-p_-}{2} = \mp p \pm p_\pm$. Equation 21 becomes $w_\pm = \int dx\ w_1(x)w_2(\mp x \pm x_\pm)$ ; $v_\pm = \int dp\ v_1(p)v_2(\mp p \pm p_\pm)$ which in terms of the convolution operation "*" can be written simply as

*Equation 22:* $w_\pm = w_1 * w_2^\pm$, $v_\pm = v_1 * v_2^\pm$ with $w_2^\pm = w_2(\pm x)$ and $v_2^\pm = v_2(\pm p)$

From the inequality [24, 32],

*Equation 23:* $e^{2H[A*B]} \geq e^{2H[A]} + e^{2H[B]}$,

and with [24]

*Equation 24:* $H[w_2^+] = H[w_2^-]$ and $H[v_2^+] = H[v_2^-]$

we get from Equation 22

*Equation 25:* $H[w_\pm] = H[w_1 * w_2^\pm] \geq \frac{1}{2}\ln\{e^{2H[w_1]} + e^{2H[w_2]}\}$; $H[v_\mp] = H[v_1 * v_2^\mp] \geq \frac{1}{2}\ln\{e^{2H[v_1]} + e^{2H[v_2]}\}$.

The preceding equation leads to state-dependent inequalities satisfied by separable pure states.

*Equation 26:* $H[w_\pm] + H[v_\mp] \geq \frac{1}{2}\ln\{(e^{2H[w_1]} + e^{2H[w_2]})(e^{2H[v_1]} + e^{2H[v_2]})\} \equiv h[w_i, v_j]$

Hence if

*Equation 27:* $H[w_\pm] + H[v_\mp] < \frac{1}{2}\ln\{(e^{2H[w_1]} + e^{2H[w_2]})(e^{2H[v_1]} + e^{2H[v_2]})\} \equiv h[w_i, v_j]$

then we have <u>entangled states</u>.

From the BBM inequality [27], we can write

*Equation 28:* $H[w_j] + H[v_j] \geq \ln\pi e$ with $j = 1,2$.

This yields from Equation 26,

*Equation 29:* $H[w_\pm] + H[v_\mp] \geq \frac{1}{2} \ln\{2(\pi e)^2 + e^{2H[w_1]+2H[v_2]} + e^{2H[w_2]+2H[v_1]}\}$.

Using Equation 28 once again, we can write Equation 29 in terms of only $v_j$ (or $w_j$) as $H[w_\pm] + H[v_\mp] \geq \frac{1}{2}\ln\{2(\pi e)^2 + (\pi e)^2[e^{2(H[v_2]-H[v_1])} + e^{-2(H[v_2]-H[v_1])}]\}$ or simply

*Equation 30:* $H[w_\pm] + H[v_\mp] \geq \frac{1}{2}\ln\{2(\pi e)^2(1 + \cosh\Delta H)\} \geq \ln 2\pi e$

where $\Delta H \equiv H[v_2] - H[v_1]$. One can compute that the minimum value of the argument of the natural logarithm function occurs when $\Delta H=0$ which then yields the minimum value $4(\pi e)^2 = (2\pi e)^2$. Substituting this value, gives the last inequality in the preceding equation. Although *Equation 30* is a weaker set of state-independent inequalities, states with

*Equation 31:* $H[w_\pm] + H[v_\mp] < \ln 2\pi e$

are still <u>entangled states</u>. However entangled states can still exist for $\ln 2\pi e \leq H[w_\pm] + H[v_\mp] < h[w_i, v_j]$ owing to the stronger inequalities in Equation 26 and Equation 27.

Let us next consider mixed states. Mixed states are more commonly encountered experimentally since a system can be in any of a number of states in varying probabilities which can occur when we prepare a similar system several times. Note that a mixed state is different from a state formed by a superposition of states (which can be a pure state). A bipartite separable mixed state is more conveniently described in terms of the density matrix $\rho$.

*Equation 32:* $\rho = \sum_m \lambda_m \rho_{1m} \otimes \rho_{2m}$, where $\rho_{im} = |\psi_{im}\rangle\langle\psi_{im}|$, $i = 1, 2$, $\lambda_m \geq 0$, $\sum_m \lambda_m = 1$.

Note that $\psi_{1m} = (\psi_1(r_1))_m$ is the mth pure state for subsystem 1 and $\psi_{2m} = (\psi_2(r_2))_m$ is the mth pure state for subsystem 2. We can write the probability distribution for $x_\pm$ and $p_\pm$ as

*Equation 33:* $w_\pm = \sum_m \lambda_m w_{m\pm}$, $v_\pm = \sum_m \lambda_m v_{m\pm}$

where $w_{m\pm}$ ($v_{m\pm}$) is the probability distribution to detect $x_\pm$ ($p_\pm$) for each pure state $\psi_{im}$.

Since the Shannon entropy is a concave function [32], we have

*Equation 34:* $H[w_\pm] = H[\sum_m \lambda_m w_{m\pm}] \geq \sum_m \lambda_m H[w_{m\pm}]$; $H[v_\mp] = H[\sum_m \lambda_m v_{m\mp}] \geq \sum_m \lambda_m H[v_{m\mp}]$.

Similarly from Equation 25, we get

*Equation 35:* $H[w_\pm] \geq H[\sum_m \lambda_m w_{m\pm}] \geq \sum_m \lambda_m \frac{1}{2}\ln\{e^{2H[w_{1m}]} + e^{2H[w_{2m}]}\}$; $H[v_\mp] \geq H[\sum_m \lambda_m v_{m\mp}] \geq \sum_m \lambda_m \frac{1}{2}\ln\{e^{2H[v_{1m}]} + e^{2H[v_{2m}]}\}$.

Hence, adding the preceding equations,

*Equation 36:* $H[w_\pm] + H[v_\mp] \geq \sum_m \lambda_m \frac{1}{2} \ln\left[\left(e^{2H[w_{1m}]} + e^{2H[w_{2m}]}\right)\left(e^{2H[v_{1m}]} + e^{2H[v_{2m}]}\right)\right] \equiv \hbar[w_{im}, v_{jm}]$

for separable states where $i, j = 1,2$. Hence for <u>entangled mixed states</u> we have,

*Equation 37:* $H[w_\pm] + H[v_\mp] < h[w_{im}, v_{jm}]$

Going back to Equation 34, we can write $H[w_\pm] + H[v_\mp] \geq \sum_m \lambda_m (H[w_{m\pm}] + H[v_{m\mp}])$ and from Equation 30, we get $H[w_\pm] + H[v_\mp] \geq \sum_m \lambda_m (\ln 2\pi e) = (\ln 2\pi e) \sum_m \lambda_m = (\ln 2\pi e) \cdot 1$ from Equation 32. Hence, $H[w_\pm] + H[v_\mp] \geq (\ln 2\pi e)$ even for separable mixed states and similarly we get the same weaker conditions for entanglement in Equation 31 for mixed states

*Equation 38:* $H[w_\pm] + H[v_\mp] < (\ln 2\pi e)$.

## 5. GUP-correction to the Shannon Entropy Quantum Entanglement Criterion

Having set the notation and discussed the Shannon Entropy Quantum Entanglement Criterion in section 4, we use the results of [33, 7] to get the GUP correction. As discussed in section 1, the GUP-modified commutation relation is now given by $[x_i, k_j] = i\delta_{ij}(1 + \beta k_i^2)$ with $i$, and $j$ as in section 4 above. From Equation 3, $p_i$ and $k_i$ are related by

*Equation 39:* $k_i = \frac{1}{\sqrt{\beta}} \tan(\sqrt{\beta}\, p_i)$.

We now have for the $k$ momentum operator, analogous to $\wp(p_+, p_-)$,

*Equation 40:* $\Phi(k_+, k_-) = \frac{1}{\sqrt{2}} \phi_1\left(\frac{k_+ + k_-}{2}\right) \phi_2\left(\frac{k_+ - k_-}{2}\right)$

with

*Equation 41:* $k_\pm = k_1 \pm k_2$ and $\phi_1 = \phi_1(k_1)$, $\phi_2 = \phi_2(k_2)$

and the corresponding Shannon entropy $H[u_\pm] = -\int_{-\infty}^{\infty} dk_\pm u_\pm(k_\pm) \ln u_\pm(k_\pm)$ where $u_\pm(k_\pm) = u_\pm(k_1, k_2)$ is the probability distribution of $k_\pm$. Hence from Equation 40,

*Equation 42:* $u_\pm(k_\pm) = \int dk_\mp |\Phi(k_+, k_-)|^2 = \frac{1}{2} \int dk_\mp |\phi_1|^2 |\phi_2|^2$.

Defining $u_i(k_i) \equiv |\phi_i(k_i)|^2$, Equation 42 becomes

*Equation 43:* $u_\pm(k_\pm) = \frac{1}{2} \int dk_\mp u_1\left(\frac{k_+ + k_-}{2}\right) u_2\left(\frac{k_+ - k_-}{2}\right)$.

From Equation 41, we define $\frac{k_+ + k_-}{2} = k_1 \equiv k$, and so we get $\frac{k_+ - k_-}{2} = \mp k \pm k_\pm$ then Equation 43 becomes $u_\pm(k_\pm) = \int dk\, u_1(k)\, u_2(\mp k \pm k_\pm)$ which in terms of the convolution operation "*" becomes $u_\pm(k_\pm) = u_1 * u_2^\pm$ with $u_2^\pm \equiv u_2(\pm k)$. Using Equation 23 and Equation 24 (with $v$ replaced by $u$), we get similar to the second part of Equation 25,

Equation 44: $H[u_\mp] = H[u_1 * u_2^\mp] \geq \frac{1}{2} \ln\{e^{2H[u_1]} + e^{2H[u_2]}\}$.

The probability distributions of the operators $k$ and $p$ are such that $u_i(k_i)dk_i = v_i(p_i)dp_i$ or $u_i(k_i) = v_i(p_i)\frac{dp_i}{dk_i}$ with $i = 1, 2$. Using Equation 39, we get

Equation 45: $u_i(k_i) = \frac{v_i(p_i)}{1+\beta k_i^2}$.

From the definition of the Shannon entropy Equation 15 and Equation 45, $H[u_i] = -\int_{-\infty}^{\infty} u_i(k_i) \ln u_i(k_i)\, dk_i = -\int_{-\infty}^{\infty} \frac{v_i(p_i)}{1+\beta k_i^2} \ln\left(\frac{v_i(p_i)}{1+\beta k_i^2}\right) dk_i =$

$-\int_{-\infty}^{\infty} \frac{v_i(p_i)}{1+\beta k_i^2} \ln v_i(p_i)\, dk_i + \int_{-\infty}^{\infty} \frac{v_i(p_i)}{1+\beta k_i^2} \ln(1+\beta k_i^2)\, dk_i = -\int_{-\infty}^{\infty} \frac{v_i(p_i)}{1+\beta k_i^2} \ln v_i(p_i)\, dk_i +$

$\int_{-\infty}^{\infty} u_i(k_i) \ln(1+\beta k_i^2)\, dk_i$. Finally, using again Equation 39 for a change in the integrating variable in the first integral yields $H[u_i] = -\int_{-p_0}^{p_0} v_i(p_i) \ln v_i(p_i)\, dp_i + \int_{-\infty}^{\infty} u_i(k_i) \ln(1+\beta k_i^2)\, dk_i$ with $p_0 \equiv \frac{\pi}{2\sqrt{\beta}}$. Hence we can write

Equation 46: $H[u_i] = H[v_i] + \eta_i(\beta)$ with $\eta_i(\beta) \equiv \int_{-\infty}^{\infty} u_i(k_i) \ln(1+\beta k_i^2)\, dk_i > 0$.

$\eta_i(\beta)$ basically gives the GUP correction. From Equation 46 and Equation 44, we get

Equation 47: $H[u_\mp] \geq \frac{1}{2} \ln\{e^{2H[v_1]+2\eta_1(\beta)} + e^{2H[v_2]+2\eta_2(\beta)}\}$.

Combining the preceding equation with the inequality for $w_\pm$ in Equation 25, we get $H[w_\pm] + H[u_\mp] \geq \frac{1}{2}\ln\{e^{2\eta_1(\beta)}e^{2H[v_1]} + e^{2\eta_2(\beta)}e^{2H[v_2]}\} + \frac{1}{2}\ln\{e^{2H[w_1]} + e^{2H[w_2]}\}$ or

Equation 48: $H[w_\pm] + H[u_\mp] \geq \frac{1}{2}\ln\{(e^{2H[w_1]} + e^{2H[w_2]})(e^{2\eta_1(\beta)}e^{2H[v_1]} + e^{2\eta_2(\beta)}e^{2H[v_2]})\} \equiv h_{GUP}(w_i, v_j)$

for separable pure states. Since from Equation 46 $\eta_i(\beta = 0) = 0$ for the non-GUP case, Equation 48 reduces to Equation 26 as expected. Hence if $H[w_\pm] + H[u_\mp] < h_{GUP}(w_i, v_j)$ then we have entangled states. Clearly since $\eta_i > 0$ in Equation 46, we see from Equation 26 and Equation 48 that $h_{GUP}(w_i, v_j) > h[w_i, v_j]$. Hence with Equation 27 and Equation 48 we have a higher upper bound for the GUP-corrected entanglement condition. It is interesting to note that the same increase in the upper bound for the GUP-corrected entanglement condition was found when using variance relations [1].

We now derive the corresponding GUP-corrected expression for Equation 30. Since the BBM inequality has been shown to still hold in the GUP framework [7], we can combine Equation 46 and Equation 28 to yield

*Equation 49:* $H[u_i] + H[w_i] \geq \ln \pi e + \eta_i(\beta)$

which yields $e^{H[u_i]+H[w_i]} \geq (\pi e)e^{\eta_i(\beta)}$ or

*Equation 50:* $e^{2(H[u_i]+H[w_i])} \geq (\pi e)^2 e^{2\eta_i(\beta)}$.

From Equation 44 and Equation 25, we get $H[w_\pm] + H[u_\mp] \geq \frac{1}{2}\ln(e^{2H[w_1]} + e^{2H[w_2]}) + \frac{1}{2}\ln\{e^{2H[u_1]} + e^{2H[u_2]}\} = \frac{1}{2}\ln\{(e^{2H[w_1]} + e^{2H[w_2]})(e^{2H[u_1]} + e^{2H[u_2]})\}$ or

*Equation 51:* $H[w_\pm] + H[u_\mp] \geq \frac{1}{2}\ln(e^{2H[w_1]+2H[u_1]} + e^{2H[w_2]+2H[u_2]} + e^{2H[w_1]+2H[u_2]} + e^{2H[w_2]+2H[u_1]})$.

Because of Equation 50, we can replace the first two terms of Equation 51 to yield

*Equation 52:* $H[w_\pm] + H[u_\mp] \geq \frac{1}{2}\ln((\pi e)^2 e^{2\eta_1(\beta)} + (\pi e)^2 e^{2\eta_2(\beta)} + e^{2H[w_1]+2H[u_2]} + e^{2H[w_2]+2H[u_1]})$.

Let us look at the third term of the preceding equation. Using Equation 49 to eliminate $H[w_1]$, we have $e^{2H[w_1]+2H[u_2]} \geq e^{2(\ln \pi e + \eta_1(\beta) - H[u_1]) + 2H[u_2]} = (\pi e)^2 e^{2\eta_1} e^{2(H[u_2] - H[u_1])}$. Similarly, we have $e^{2H[w_2]+2H[u_1]} \geq (\pi e)^2 e^{2\eta_2} e^{-2(H[u_2]-H[u_1])}$. We can then replace the last 2 terms in Equation 52 to give $H[w_\pm] + H[u_\mp] \geq \frac{1}{2}\ln((\pi e)^2 e^{2\eta_1(\beta)} + (\pi e)^2 e^{2\eta_2(\beta)} + (\pi e)^2 e^{2\eta_1} e^{2(H[u_2]-H[u_1])} + (\pi e)^2 e^{2\eta_2} e^{-2(H[u_2]-H[u_1])})$. The right hand side readily simplifies to

*Equation 53:* $H[w_\pm] + H[u_\mp] \geq \ln \pi e + \frac{1}{2}\ln\{(e^{2\eta_1(\beta)} + e^{2\eta_2(\beta)}) + (e^{2\eta_1} e^{2\Delta H_{GUP}} + e^{2\eta_2} e^{-2\Delta H_{GUP}})\}$

with $\Delta H_{GUP} \equiv H[u_2] - H[u_1]$. As in section 4, Equation 30, let us attempt to rewrite the inequality independent of the Shannon entropy difference $\Delta H_{GUP}$. We can solve for the minimum of $e^{2\eta_1(\beta)} + e^{2\eta_2(\beta)} + e^{2\eta_1} e^{2\Delta H_{GUP}} + e^{2\eta_2} e^{-2\Delta H_{GUP}}$. It occurs at $\Delta H_{GUP} = \frac{1}{2}\ln\left(\frac{e^{2\eta_2(\beta)}}{e^{2\eta_1(\beta)}}\right)$. This yields a minimum value of $e^{2\eta_1(\beta)} + e^{2\eta_2(\beta)} + 2e^{\eta_1(\beta)}e^{\eta_2(\beta)} = (e^{\eta_1(\beta)} + e^{\eta_2(\beta)})^2$. We can substitute this into the argument of the natural logarithm function in Equation 53 to get $H[w_\pm] + H[u_\mp] \geq \ln \pi e + \ln(e^{\eta_1(\beta)} + e^{\eta_2(\beta)})$ or

*Equation 54:* $H[w_\pm] + H[u_\mp] \geq \ln 2\pi e + \ln\left(\frac{e^{\eta_1(\beta)} + e^{\eta_2(\beta)}}{2}\right)$.

Equation 54 is the GUP-corrected version of Equation 30. Unlike Equation 30, the preceding equation is not state-independent. The GUP-corrected version of Equation 31 for entangled states is then

*Equation 55:* $H[w_\pm] + H[u_\mp] < \ln 2\pi e + \ln\left(\frac{e^{\eta_1(\beta)} + e^{\eta_2(\beta)}}{2}\right).$

As before the upper bound for entangled states is increased when compared to Equation 31.

The GUP correction for mixed states expressed in Equation 32 follows naturally. Similar to Equation 33,

*Equation 56:* $w_\pm = \sum_m \lambda_m w_{m\pm}, \; u_\pm = \sum_m \lambda_m u_{m\pm}.$

where $u_{m\pm}$ is the probability distribution to detect $k_\pm$ for each pure state $\psi_{im}$.

Similar to Equation 46,

*Equation 57:* $H[u_{im}] = H[v_{im}] + \eta_{im}(\beta)$ with $\eta_{im}(\beta) \equiv \int_{-\infty}^{\infty} u_{im}(k_i) \ln(1 + \beta k_i^2) \, dk_i > 0$ with $\eta_{im}(\beta), \; i = 1,2$

giving the GUP correction. Using the concavity of the Shannon entropy function, Equation 56 gives

*Equation 58:* $H[u_\mp] = H[\sum_m \lambda_m u_{m\mp}] \geq \sum_m \lambda_m H[u_{m\mp}].$

Similar to Equation 47, $H[u_{m\mp}] \geq \frac{1}{2}\ln\{e^{2H[v_{1m}] + 2\eta_{1m}(\beta)} + e^{2H[v_{2m}] + 2\eta_{2m}(\beta)}\}$. Hence Equation 58 becomes $H[u_\mp] \geq \sum_m \lambda_m \frac{1}{2}\ln\{e^{2H[v_{1m}] + 2\eta_{1m}(\beta)} + e^{2H[v_{2m}] + 2\eta_{2m}(\beta)}\}$. Adding the preceding equation to Equation 35,

*Equation 59:* $H[w_\pm] + H[u_\mp] \geq \sum_m \lambda_m \frac{1}{2} \ln\{(e^{2H[w_{1m}]} + e^{2H[w_{2m}]})(e^{2H[v_{1m}] + 2\eta_{1m}(\beta)} + e^{2H[v_{2m}] + 2\eta_{2m}(\beta)})\} \equiv \hbar_{GUP}(w_{im}, v_{jm})$

for separable states and hence for entangled mixed states, we have $H[w_\pm] + H[u_\mp] < \hbar_{GUP}(w_{im}, v_{jm})$. Hence similar to the conclusion from Equation 48, we get a higher upper bound for the GUP-corrected entanglement condition as compared to Equation 37 since $\hbar_{GUP}(w_{im}, v_{jm}) > \mathrm{h}_{GUP}(w_{im}, v_{jm})$ as in the pure state case.

Returning to Equation 34 and Equation 58, and similar to Equation 54, $H[w_\pm] + H[u_\mp] \geq \sum_m \lambda_m (H[w_{m\pm}] + H[u_{m\mp}]) \geq \sum_m \lambda_m \left(\ln 2\pi e + \ln\left[\frac{e^{\eta_{1m}(\beta)} + e^{\eta_{2m}(\beta)}}{2}\right]\right) = \sum_m \lambda_m (\ln 2\pi e) + \sum_m \lambda_m \ln\left[\frac{e^{\eta_{1m}(\beta)} + e^{\eta_{2m}(\beta)}}{2}\right]$ and from Equation 32, $H[w_\pm] + H[u_\mp] \geq \ln 2\pi e + \sum_m \lambda_m \ln\left[\frac{e^{\eta_{1m}(\beta)} + e^{\eta_{2m}(\beta)}}{2}\right]$. The entanglement criterion is then

*Equation 60:* $H[w_\pm] + H[u_\mp] < \ln 2\pi e + \sum_m \lambda_m \ln\left[\frac{e^{\eta_{1m}(\beta)} + e^{\eta_{2m}(\beta)}}{2}\right].$

Equation 60 is the GUP-corrected expression for Equation 38 for mixed states which shows an increase in the upper bound of Equation 38.

# 6. Conclusions

The study of the effects of the minimal length and the generalized uncertainty principle on entanglement involving continuous variable systems has been confined to the variance form of the HUP [1]. Since the more modern approach of expressing the uncertainty principle is through the use of entropic measures [34], as discussed in section 3 above, we conducted a study of how the entanglement criterion using the Shannon entropic uncertainty relation is modified by the generalized uncertainty principle. As expected, the upper bounds of the entanglement criteria for pure and mixed states in bipartite systems are increased just like in the variance form [1] from which we can infer as in [1] that the increase in the bounds due to GUP effects enhances entanglement making the quantum effects more pronounced. Intuitively, we expect the introduction of gravitational effects to increase the entropy and hence the uncertainty leading to a higher upper bound for entanglement. It should also be noted that the main difference between the weaker inequality entanglement criteria using the Shannon entropy for pure and mixed states without the GUP (Equation 31 and Equation 38) and with the GUP (Equation 55 and Equation 60) is the state-dependence due to the GUP correction $\eta_i(\beta)$ and $\eta_{im}(\beta)$ defined in Equation 46 and Equation 57. However, a state-independent entanglement criterion can apparently be derived using the Renyi and Tsallis entropies since state-independent entropic uncertainty relations with GUP have been obtained in [33]. Since the entropic bounds are increased we expect also an increase in the entanglement criteria upper bounds.

The connection of gravity and entanglement via the entropic uncertainty relations may not be surprising after all. Recently the idea of gravity as an entropic force [35] has been proposed. Gravity can then be thought of as resulting from the tendency of physical systems to increase their entropy. Furthermore, it has been proposed [36] that the entanglement (measured in terms of entanglement von Neumann entropy) of degrees of freedom in quantum systems can give rise to the emergence of spacetime in the gravity picture from field theory with the gravity- gauge theory correspondence (AdS-CFT) [37-39]. Whereas these studies indicate the effect of quantum entanglement on gravity, our work opens up yet another avenue of the effect of gravity on quantum entanglement. Gravity and quantum entanglement seem to be mutually related.

As mentioned above, with the GUP correction obtained for the Renyi and Tsallis entropic uncertainty relations in [33], it will be interesting to study carefully the GUP effect on the entanglement criteria based on the Renyi entropy as a generalization of the Shannon entropy. Preliminary work on the GUP effects on the Renyi entropic uncertainty relations has indeed shown an increase in the entanglement criteria upper bound [40].